\documentclass[%
 reprint,
 amsmath,amssymb,
 aps,pra,twocolumnn,superscriptaddress,
]{revtex4-2}

\usepackage{dcolumn}
\usepackage{bm}
\usepackage[final]{graphicx}
\usepackage{times,bbm,amsmath,amssymb}
\usepackage{epsfig,color}
\usepackage{hyperref}
\usepackage{float,siunitx}
\usepackage[caption = false]{subfig}
\usepackage[english]{babel}
\usepackage{thumbpdf,enumerate}
\usepackage{booktabs}
\usepackage{sidecap}
\usepackage[scaled=.8]{couriers}
\usepackage{pstricks}
\usepackage{multirow}
\usepackage{placeins}
\usepackage{pst-grad}
\usepackage{epigraph}
\usepackage{longtable}
\usepackage{booktabs}
\usepackage{multirow}
\usepackage{gensymb}
\usepackage{array}
\usepackage{soul}

\def\ket#1{ \vert #1 \rangle}

\def\Tr{\text{Tr}}

\begin{document}

\preprint{APS/123-QED}

\title{Experimental witnessing of the quantum channel capacity in the presence of correlated noise }

\author{V. Cimini}
\affiliation{Dipartimento di Scienze, Universit\`a degli Studi Roma Tre, Via della Vasca Navale 84, 00146, Rome, Italy}
\author{I. Gianani}
\affiliation{Dipartimento di Scienze, Universit\`a degli Studi Roma Tre, Via della Vasca Navale 84, 00146, Rome, Italy}
\affiliation{Dipartimento di Fisica, Sapienza Universit\`a di Roma, Piazzale Aldo Moro 5, 00185, Rome, Italy}
\author{M.F. Sacchi}
\affiliation{Istituto di Fotonica e Nanotecnologie - CNR, Piazza Leonardo Da Vinci 32, 20133 Milan, Italy}
\affiliation{QUIT group, Dipartimento di Fisica, Universit\`a degli Studi di Pavia, Via A. Bassi 6, 27100 Pavia, Italy}
\author{C. Macchiavello}
\affiliation{QUIT group, Dipartimento di Fisica, Universit\`a degli Studi di Pavia, Via A. Bassi 6, 27100 Pavia, Italy}
\affiliation{INFN Sezione di Pavia, via Bassi 6, I-27100, Pavia, Italy}
\affiliation{Istituto Nazionale di Ottica - CNR, Largo Enrico Fermi 6, 50125 Florence, Italy}
\author{M. Barbieri}
\affiliation{Dipartimento di Scienze, Universit\`a degli Studi Roma Tre, Via della Vasca Navale 84, 00146, Rome, Italy}
\affiliation{Istituto Nazionale di Ottica - CNR, Largo Enrico Fermi 6, 50125 Florence, Italy}

\email{}

\begin{abstract}
We present an experimental method to detect lower bounds to the
quantum capacity of two-qubit communication channels.  We consider an
implementation with polarisation degrees of freedom of two photons and
report on the efficiency of such a method in the presence of
correlated noise for varying values of the correlation strength.  The
procedure is based on the generation of separable states of two qubits
and local measurements at the output. We also compare the performance
of the correlated two-qubit channel with the single-qubit channels
corresponding to the partial trace on each of the subsystems, thus
showing the beneficial effect of properly taking into account correlations to 
achieve  a larger quantum capacity.
\end{abstract}

\maketitle


\section{Introduction}
Quantum communication channels in the presence of correlations among
subsequent uses have attracted much attention recently.  Correlated
qubit channels were originally investigated in the context of
classical information transmission, showing that for certain ranges of
the correlation strengths the generation of entanglement among
subsequent uses is beneficial to enhance the amount of transmitted
information \cite{memory}.  New interesting features then emerged in
the study of quantum memory (or correlated) channels by modeling of
relevant physical examples, including depolarizing
channels~\cite{MMM}, Pauli channels~\cite{mpv04,daems,dc}, dephasing
channels~\cite{hamada,dbf,ps,gabriela,lidar}, amplitude damping
channels \cite{vsdamp}, Gaussian channels~\cite{cerf},
lossy bosonic channels~\cite{mancini,lupo}, spin chains~\cite{spins},
collision models~\cite{collision}, and a micromaser
model~\cite{micromaser} (for a recent review on quantum channels with
memory effects see Ref.~\cite{memo_review}).

Quantum channels can be characterised completely by means of quantum
process tomography~\cite{chuang}, a well established technique that
requires a number of measurement settings (in an entanglement-based
scenario, or otherwise a number of measurement settings times number
of state preparations in a single system scenario) that scales as
$d^4$, where $d$ is the arbitrary finite dimension of the quantum
system which is sent through the communication
channel~\cite{lobino,sudafrica,quantum}.

Less expensive procedures, with a number of measurement settings
scaling as $d^2$, have been recently proposed to detect specific
properties of a quantum channel that do not need a complete
characterisation, such as for example its entanglement breaking
property \cite{mapdet} or its non-Markovian character \cite{noma-wit}.
A central feature to quantify the channel ability to convey
information is the channel capacity. Efficient procedures have been
recently proposed to detect lower bounds to the capacity of an unknown
quantum communication channel that avoid quantum process tomography,
in particular for the quantum capacity \cite{qcap-det} and the
classical capacity \cite{ms19}. The performance of the procedure
proposed in Ref. \cite{qcap-det} was demonstrated experimentally for
single qubit channels in \cite{exp}.

In the present paper we demonstrate experimentally that channel capacity witnesses can capture correlations among multiple quantum communication channels.  
The procedure originally proposed in Ref. \cite{qcap-det} efficiently detects
lower bounds to the quantum capacity of correlated two-qubit channels, which we
compare with the theoretical values reported in Ref. \cite{qcap-corr}. The two-qubit correlated channels are
implemented by acting with liquid crystals affecting the polarisation of two photons. The correlation level is set by controlling the relative operation conditions of the two liquid crystals. The witnessing procedure, that works
for unknown channels, is demonstrated without the need of generating
entangled states. 
\section{Capacity witness}
We briefly review the general method introduced in
Ref. \cite{qcap-det} to experimentally achieve lower bounds to the
quantum capacity of noisy channels by few local measurements. This technique has been introduced in order to reduce the experimental requirements on channel characterization. It is part of an ongoing effort to make state \cite{entwitness1, entwitness2,entwitness3, entwitness4, entwitness5, coherence1, coherence2, coherence3, nonclassicality1, nonclassicality2, compressed_tomo} and process \cite{prl_luise, Teo_tomography, cho_tomo} reconstruction more efficient. In addition, calculating the channel capacity demands assessing infinite uses of the channel, a task which can not always be carried out analytically. This also motivates the search of more practical bounds on the capacity.

The method can adopt a fixed maximally bipartite entangled state of two
copies of the system, where just one copy enters the quantum channel
and suitable separable measurements are jointly performed on the
output copy and the second untouched reference copy. Equivalently, the
method can also be carried out by suitable preparation of different
ensembles of a single copy at the input of the channel, with
corresponding output measurements. Since in the present experimental
implementation the second option is followed, we will specifically
focus on this second scenario. We observe that in both strategies the number of measurements is less than the one for the process tomography of the channel~\cite{processresources}, albeit the separable-state case requires more measurements. On the other hand, this alleviates the difficulty of generating multiple entangled pairs at once.

Let us denote the action of a generic memoryless quantum channel on a
single system as ${\cal E}$. The quantum
capacity $Q$ is measured in qubits per channel use and is defined as
\cite{lloyd,barnum,devetak} $Q=\lim _{N\to \infty}\frac {Q_N}{N}$,
where $Q_N = \max _{\rho } I_c (\rho , {\cal E}^{\otimes  N})$, and $I_c(\rho ,
{\cal E})$ denotes the coherent information \cite{schumachernielsen}
\begin{eqnarray} I_c(\rho , {\cal E}) = S[{\cal E} (\rho )] - S_e
(\rho, {\cal E})\;.\label{ic} \end{eqnarray} 
In Eq. (\ref{ic}),
$S(\rho )=-\Tr [\rho \log _2 \rho ]$ is the von Neumann entropy and
$S_e (\rho, {\cal E})$ represents the entropy exchange \cite{schumacher}, i.e.
$S_e (\rho, {\cal E})= S[({\cal I}_R \otimes {\cal
  E})(|\Psi _\rho \rangle \langle \Psi _\rho |)] $,
where $|\Psi _\rho \rangle $ is any purification of $\rho $ by means of a 
reference quantum system $R$, namely 
$\rho =\Tr _R [|\Psi _\rho \rangle \langle \Psi _\rho|]$.

We recall that for any complete set of orthogonal projectors
$\{\Pi _i\}$ one has \cite{NC00} $S(\rho )\leq S(\sum _i \Pi _i \rho
\Pi _i)$. It follows that from any orthonormal basis $\{ |\Phi _i
\rangle \}$ for the tensor product of the reference and the system
Hilbert spaces one obtains the following bound to the entropy exchange
\begin{eqnarray}
S_e\left (\rho , {\cal E} \right )\leq H (\vec p)\;,  
\label{se-bound}
\end{eqnarray}
where $H(\vec p)$ denotes the Shannon entropy $H(\vec p)=-\sum _{i}
p_i \log _2 p_i$ for the vector of 
probabilities $\{p_i\}$,  with 
\begin{eqnarray}
p_i = \Tr [( {\cal I}_R \otimes {\cal E} ) (|\Psi _\rho \rangle
  \langle \Psi _\rho |) |\Phi _i \rangle\langle\Phi _i|] \;.
\label{pimeas}
\end{eqnarray}
Therefore, from Eq. (\ref{se-bound}) it follows that for any $\rho$ and $\vec p$ one has 
the chain of bounds 
\begin{eqnarray}
Q \geq Q_1 \geq I_c(\rho , {\cal E})\geq S\left [{\cal E} (\rho )\right ]-H(\vec p) 
\equiv Q_{DET}
\;.\label{qvec}
\end{eqnarray}
A capacity witness $Q_{DET}$ for the quantum capacity $Q$ can then be
accessed without requiring full process tomography of the quantum
channel as long as the entropy of the output state of the system and a set of probabilities
$\{p_i\}$ as in Eq. (\ref{pimeas}) are experimentally measured.

The experimental measurement of $Q_{DET}$ can then be performed, based
on a maximally entangled state as the input~\cite{exp}. We consider a
complete set of observables $\{X_i \}$ for the space of system
operators, and the maximally entangled state
$\ket{\phi^+}=\frac{1}{\sqrt d}\sum_{k=0}^{d-1}\ket{k}\ket{k}$, with
respect to the bipartite space $\mathcal{H}_R \otimes \mathcal{H}$,
with $d=\mbox{dim}(\mathcal{H})=\mbox{dim}(\mathcal{H}_R)$.  By
comparison with Eq. (\ref{pimeas}), with the identification $|\Psi
_\rho \rangle = |\phi ^+ \rangle $ (i.e. $\rho = I /d$), the
input/output correlations allow to reconstruct probability vectors
$\vec p$ for all possible inequivalent bipartite orthonormal bases $\{
|\Phi _i \rangle \}$ that can be spanned by the set of measured
observables $\{
X_i ^\tau \otimes X_i \}$, where $\tau $ denotes the transposition
operation. The detection method is then supplemented by classical
optimization over all such possible bases. Moreover, the measurement
setting with observables $\{ X_i\}$ clearly allows to reconstruct
${\cal E}(I/d)$, and then to evaluate the entropy contribution
$S[{\cal E}(I/d)]$.

Alternatively, one can devise a detection method that does not require
initial entanglement,  and thus an additional reference system. Indeed,
one can easily verify the identity \cite{bellob}
\begin{eqnarray}
&&\langle X_i ^\tau  \otimes X_i \rangle \equiv
\Tr [( {\cal I}_R \otimes {\cal E} )(|\phi ^+ \rangle \langle \phi ^+ |) 
  ( X_i^\tau  \otimes X_i )]\nonumber \\& &
=\frac 1 d 
\Tr[X_i {\cal E}(X_i)]\;,
\label{eq:bellob}
\end{eqnarray}
Then, the expectation values $\langle X_i ^\tau \otimes X_i \rangle $
can be reconstructed by preparing the system in the eigenstates of
$X_i$, and measuring $X_i$ at the output of the channel, without need
of using entangled input states. These still give access to the
probabilities $p_i$ in \eqref{pimeas} with a classical optimisation,
as described before.

In the case of single-qubit channels, a measurement setting based on
the customary Pauli operators $\{\sigma _X, \sigma _Y, \sigma _Z \}$
provides probability vectors pertaining to the following inequivalent
bases \cite{qcap-det}
\begin{eqnarray}
B_{1}
=&\{a|\Phi^{+}\rangle+b|\Phi^{-}\rangle,-b|\Phi^{+}\rangle+a|\Phi^{-}\rangle,\label{b1}\\
&c|\Psi^{+}\rangle+d|\Psi^{-}\rangle,-d|\Psi^{+}\rangle+c|\Psi^{-}\rangle\};\nonumber\\
B_{2}
=&\{a|\Phi^{+}\rangle+b|\Psi^{+}\rangle,-b|\Phi^{+}\rangle+a|\Psi^{+}\rangle,\label{b2}\\
&c|\Phi^{-}\rangle+d|\Psi^{-}\rangle,-d|\Phi^{-}\rangle+c|\Psi^{-}\rangle\};\nonumber\\
B_{3}
=&\{a|\Phi^{+}\rangle+ib|\Psi^{-}\rangle,ib|\Phi^{+}\rangle+a|\Psi^{-}\rangle,\label{b3}\\
&c|\Phi^{-}\rangle+id|\Psi^{+}\rangle,id|\Phi^{-}\rangle+c|\Psi^{+}\rangle\};\nonumber 
\end{eqnarray} 
where $| \Phi ^\pm \rangle ={1}/{\sqrt 2}(|00 \rangle \pm |11 \rangle
)$ and $|\Psi ^\pm \rangle ={1}/{\sqrt 2}(|01 \rangle \pm |10 \rangle
)$ denote the Bell states, and $a,b,c,d$ are real numbers such that
$a^2+b^2=c^2+d^2=1$.  After collecting the measurement outcomes, the
capacity witness $Q_{DET}$ is then maximized over the three bases
$B_1,B_2,B_3$, and by varying the independent parameters $b$ and $d$,
namely
\begin{eqnarray}
Q_{DET}&=&\max _{j=1,2,3}\max _{b,d}Q_{DET}(B_j,b,d) \nonumber \\
&=&S[\mathcal{E}(I/2)]-\min _{j=1,2,3}\min _{b,d}
H[\vec{p}(B_j,b,d)]\;,
\end{eqnarray}
where for each $j$, the $i$-th component of the four-dimensional
probability vector $\vec p (B_j,b,d)$ corresponds to
Eq. (\ref{pimeas}), where $|\Phi _i \rangle $ is one of the four
states in the basis $B_j$. As detailed above, the input entangled
state can be replaced with the set of eigenvectors of
$\sigma_X,\sigma_Y$ and $\sigma_Z$, leading to an equivalent
reconstruction.

For a two-qubit channel, as in the present experimental
implementation, the set of observables is chosen as $\{ \sigma _i
\otimes \sigma _j \}$, with $i,j=X,Y,Z$. The input/output correlation
allows to obtain probability vectors $\vec p(B_j,b_j,d_j;B_l,b_l,d_l)$
with $16$ elements, corresponding to the bases obtained by the tensor
product of $B_j$ and $B_l$. The optimization of the capacity witness
is then obtained by maximisation over 9 bases, each of them
continuously parametrized by 4 independent real variables.
\section{Experiment}
In our experiment, we consider a correlated two-qubit unital channel,
where a Pauli $X$ operation acts on each qubit with a certain
probability $p$, jointly or separately thus defining a degree of
correlation $\mu$. More specifically, the Kraus decomposition of the
channel takes the form
\begin{equation}
\label{eq:canale}
    \mathcal{E}_2(\rho) = \sum_{i_1,i_2=0,X} A_{i_1,i_2}\sigma_{i1}\otimes\sigma_{i2}\rho \sigma_{i1}\otimes\sigma_{i2}, 
\end{equation}
where $\sigma_0$ is the identity and all coefficients of the Pauli
operations other than $A_{0,0}$, $A_{0,X}$, $A_{X,0}$, $A_{X,X}$
vanish. In the above form we have $A_{0,X}=A_{X,0}$, and we can
express the action of the channel in terms of the two parameters
\begin{equation}
    p=1-A_{0,0}-A_{X,0},
\end{equation}
and 
\begin{equation}
\mu=1-\frac{A_{X,0}}{p(1-p)}.
\end{equation}
The above two-qubit channel is unitarily equivalent to a correlated
dephasing channel and the quantum capacity is known to be \cite{dbf,ps}
\begin{equation}
\begin{aligned}
\label{qcapvera}
    Q&=2-pH_2[(1-p)(1-\mu)]\\
     &-(1-p)H_2[p(1-\mu)]-H_2(p),
    \end{aligned}
\end{equation}
where $H_2(p)$ denotes the binary Shannon entropy. For this case, the
capacity witness $Q_{DET}$ is expected to provide a strict bound to
the actual capacity~\cite{qcap-corr}.

Notice that the channel acts locally on each single qubit, up to a
unitary operation, as a dephasing operation $\mathcal{E}_1(\rho_1)=p
\rho_1+(1-p)\sigma_X\rho_1\sigma_X$, independently of the value of
$\mu$.  If the two single-qubit channels are independent, their
combined capacity could simply be found by setting $\mu=0$
in~\eqref{qcapvera}; also in this case, the detectable bound on the
capacity is tight.
\begin{figure}[htb]
    \includegraphics[width=\columnwidth]{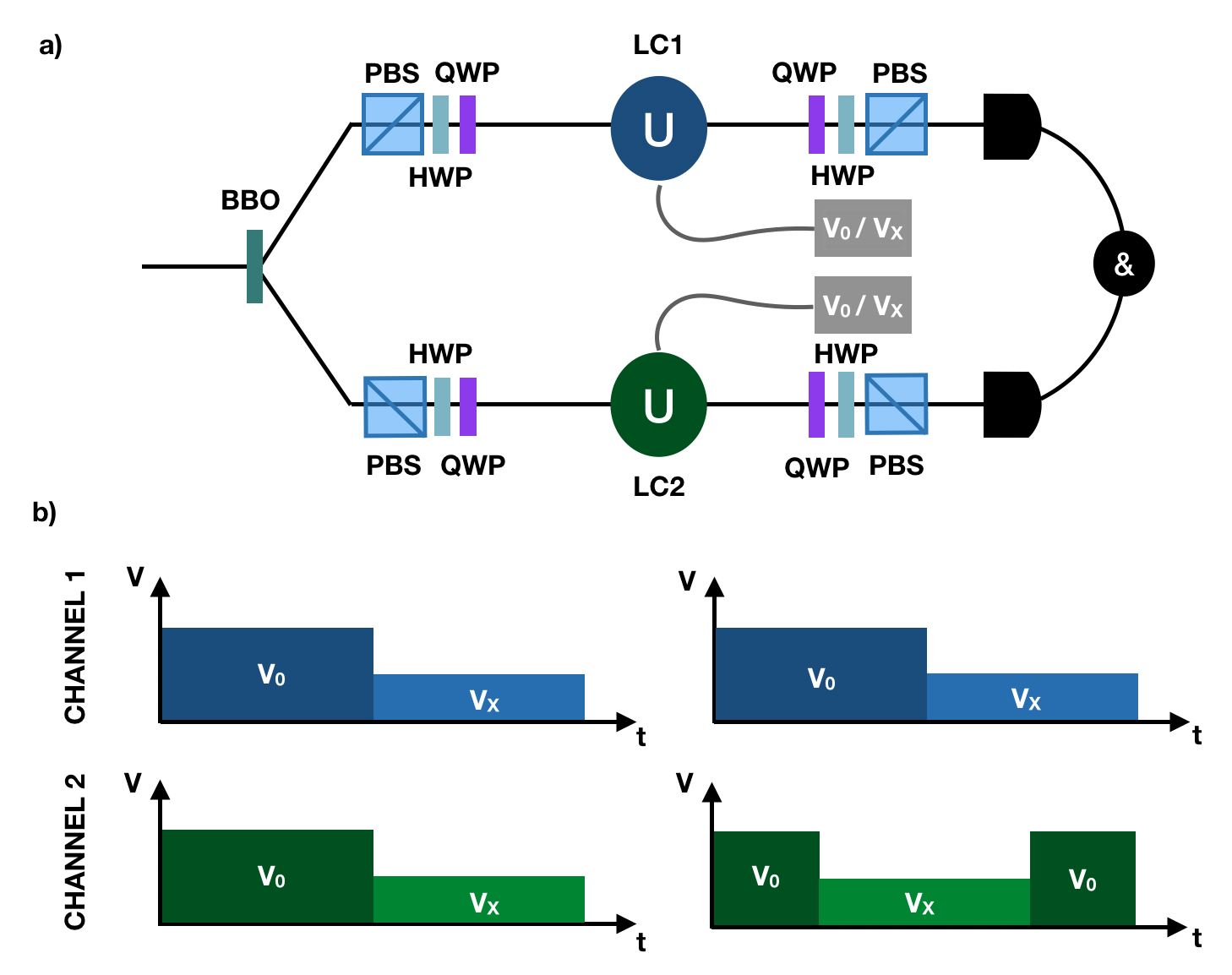}
    \caption{{\it The experiment.} a. The setup adopts a nonlinear
      crystal to generate two-photon states. These are then prepared
      in the quorum of polarisation states for the capacity witness by
      means of a polarising beam splitter (PBS), a quarter-wave plate
      (QWP), and a half-wave plate (HWP). The channel
      \eqref{eq:canale} is implemented by a pair of liquid crystal
      (LC) elements subject to time-varying voltage levels $V_0$ and
      $V_X$. Polarisation measurements are carried out by a sequence
      of QWP-HWP-PBS and single-photon detection. Coincidence
      measurements are then performed. b. Voltage sequences applied on
      the two channels for $p=1/2$, in the perfectly correlated
      $\mu=1$ (left) and uncorrelated $\mu=0$ (right) cases. }
    \label{fig:setup}
\end{figure}
\par The channels’ effect on the photon statistics is simulated using mixtures of operations on the polarisation of single photons. This is for our purposes equivalent to test the witnessing method after a direct implementation of the channel. Photon pairs are produced by parametric down
conversion source (CW-pumped at $\lambda_p=405$nm, degenerate type-I
emission at $\lambda=2\lambda_p=810$nm with 7.5nm-bandwidth filters, see Fig.
\ref{fig:setup}a). The active elements are liquid crystal plates, 
whose birefringence can be varied by applying a voltage. We thus set
two different levels for the voltage, namely $V_0$, corresponding to
the identity, and $V_X$, corresponding to the Pauli-X, for different
times $t_0$ and $t_X$, respectively thus defining $p$. 

The key to introducing correlations between the two channels is the control of the relative timings of the LCs. Consider, for instance, the case for $p=1/2$: during the total counting time $T_c=8$s, on each channel the LCs remain, overall, at $V_X$ for $t_X=4$s and at $V_0$ for $t_0=4$s. In the first arm, we simply switch between the two voltage levels halfway during the measurement (Fig. \ref{fig:setup}b). The two channels will be maximally correlated, $\mu=1$, if we change settings of the LC in the second arm at exactly the same time (Fig. \ref{fig:setup}b); on the opposite extreme, the channels act independently, $\mu=0$ if the four possible settings $(V_X,V_X)$, $(V_X,V_0)$, $(V_0,V_X)$, and $(V_0,V_0)$ all occur for same duration (Fig. \ref{fig:setup}b). We can access intermediate values of $\mu$ by anticipating the switching time from $V_0$ to $V_X$ in channel two, ensuring it is switched back again to maintain an equal amount of time for both settings; this also guarantees that $A_{0,X}=A_{X,0}$. The same reasoning can be applied to other values of $p$ and $\mu$, following the prescriptions detailed in Table 1 in the Appendix. Our implementation of the channels is a simple one, and has the advantage of providing good control of the level of correlations. However, we can rely on such realization for our goal of testing the channel capacity witness. 


In order to avoid recurring to four-qubits entangled states, the
capacity witness has been measured by using the separable-input
strategy described in the previous section. In our scheme, we encode
the eigenvalues of $\sigma_Z$ as the horizontal $\vert H \rangle$ and
vertical $\vert V \rangle$ polarisations; the eigenvalues of
$\sigma_X$ as $\vert D \rangle =(\vert H \rangle + \vert V \rangle
)/\sqrt{2}$, and $\vert A \rangle =(\vert H \rangle + \vert V \rangle
)/\sqrt{2}$; the eigenvalues of $\sigma_Y$ as $\vert L \rangle =(\vert
H \rangle +i \vert V \rangle )/\sqrt{2}$, and $\vert R \rangle =(\vert
H \rangle -i \vert V \rangle )/\sqrt{2}$. All these states can be
prepared and measured by a suitable combination of half- and
quarter-wave plates~\cite{tomography}. All relevant probabilities are
then evaluated based on coincidence count rates; no correction for
accidental events and  dark counts has been implemented.

When estimating the probabilities in \eqref{pimeas}, experimental
imperfections may lead to small negative values. These are well known
artifacts that may occur also in quantum
tomography~\cite{tomography}. When these are used in the expression of
the entropy, they lead to imaginary values; we found that just considering
the real part provides a sufficient regularisation.




We use the single-qubit and two-qubit witness for the
combined capacity of the two channels $Q_{\rm tot}$, as well as the
capacities of the individual channels $Q_1$ and $Q_2$. As these bounds
are known to be tight, we can adopt $Q_{\rm lim}{=}Q_1+Q_2$ as the
capacity for the independent use of the channels, namely without
exploiting the presence of correlations. 
Therefore, as far as the channel is modelled by Eq. (\ref{eq:canale}), we can assess whether the channels present correlations based on the experimental data. Without any assumption on the form of the channel, the experimental data can only show that the joint use of the channels (whatever they are) provides better bounds to the quantum capacity.
\begin{figure}[htb]
    \includegraphics[width=\columnwidth]{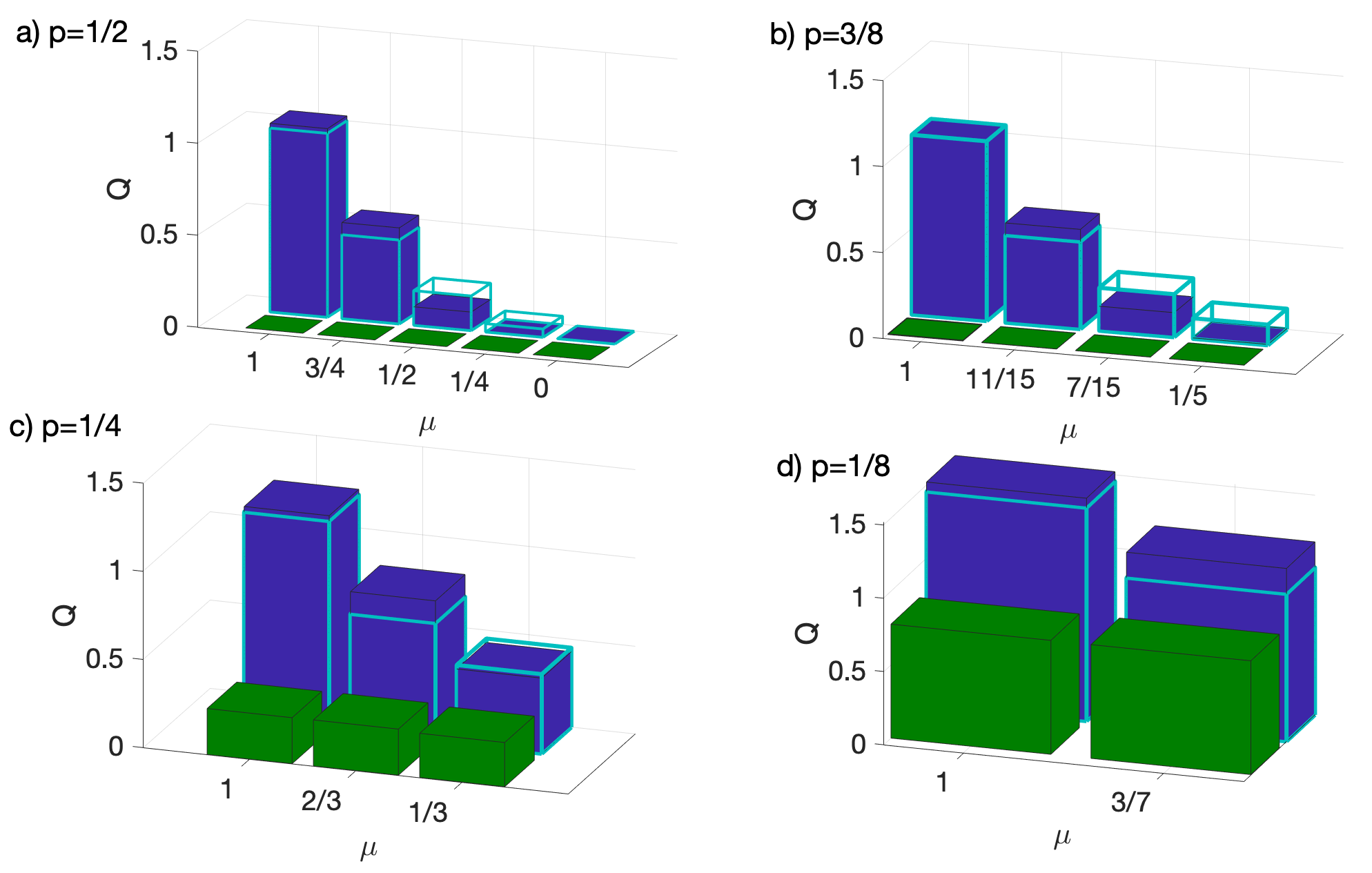}
    \caption{{\it Experimental bounds on the quantum capacity.} In all
      panels, the blue columns represent the experimental values of
      $Q_{\rm tot}$, to be compared to the green columns indicating
      the limit $Q_{\rm lim}$ for an independent use of the channels,
      when ignoring correlations. The cyan edges are the
      theoretical predictions for the ideal channel \eqref{eq:canale}
      given by Eq. \eqref{qcapvera}. Errors are of the order of 0.005,
      and hence are not visible on this scale.}
    \label{fig:eskimo}
\end{figure}
\par Our experimental results are
depicted in Fig. \ref{fig:eskimo} for the different values of $p$ and
$\mu$ considered in our experiment. Whenever the experimental
imperfections force a negative lower bound to the capacity, this is
replaced with zero. Some discrepancies with respect to the theoretical
expectations can be appreciated, mostly due to the fact that the LCs
do not implement the operations $\sigma_0$ and $\sigma_X$
exactly. Appendix 1 reports more experimental details. A direct comparison between theoretical and experimental bounds for $Q_{tot}$ shows that one can not be used as a limit for the other. 
\section{Discussion}
As we can see from the results reported in Fig. \ref{fig:eskimo}, the
green columns refer to the witness for the total quantum capacity of
the local channels, that corresponds to the theoretical values
reported in Eq. (\ref{qcapvera}) for $\mu=0$. It is clear from the
theoretical expression that for fixed value of $p$ the capacity of the
correlated channel is lower bounded by the value of the total capacity
of the local channels, and in particular it is an increasing function
of $\mu$ at fixed $p$.  If the channel that we are observing is
guaranteed to be of the form (\ref{eq:canale}), the detection of a
capacity larger than the corresponding theoretical value for $\mu=0$
signals the presence of correlations in the channel. This
behaviour can be qualitatively identified also in the results reported
in Fig. \ref{fig:eskimo}, where it is apparent that the blue columns
get closer to the green ones for decreasing values of $\mu$. We want
to stress, however, that the detection method that we implemented
works for any form of channel. In realistic experimental scenarios, a noisy channel will present deviations from its expected model, and in the extreme case the noise could even be completely unknown. Predictions based on a model can give useful indications but would fail at giving a reliable knowledge. On the contrary, the presented method certifies a lower bound to the quantum channel capacity by means of the only experimental data. The major advantage of our method is that it does not need a complete experimental reconstruction of the channel. In fact, this would require full process tomography and would then be much more demanding in terms of measurements required.
\section*{Acknowledgment}
The authors thank Fabio Sciarrino for the loan of scientific
equipment, and Paolo Mataloni and him for discussion.

I.G. is supported by Ministero dell’Istruzione, dell’Universit\`a e
della Ricerca Grant of Excellence Departments (ARTICOLO 1, COMMI
314-337 LEGGE 232/2016). C.M. acknowledges support by the Quantera
project QuICHE.

\newpage
\section*{Appendix}
\begin{table*}[ht]
\begin{center}
\setlength{\tabcolsep}{1.5em}
\renewcommand{\arraystretch}{2.5}
\begin{tabular}{|c | c || c | c | c | c || c |}
     \hline
    $p$ & $\mu$  & $A_{0,0}$ & $A_{0,X}$ & $A_{X,0}$ & $A_{X,X}$ & Voltage\\
   \hline
   \hline
    \multirow{5}{*}{1/2}& 0 &1/4 & 1/4 & 1/4 & 1/4 &
    \raisebox{-0.5\totalheight}{\includegraphics[width=0.2\linewidth, height=18mm]{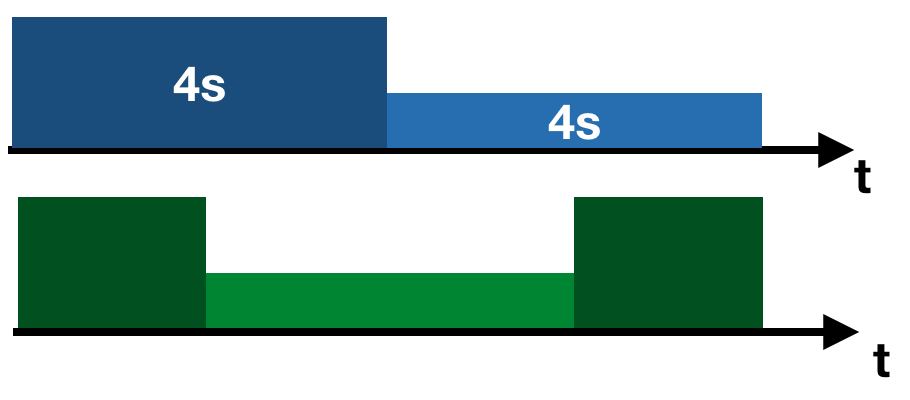}}\\
     \cline{2-7}
    & 1/4 &5/16 & 3/16 & 3/16 & 5/16 &\raisebox{-0.5\totalheight}{\includegraphics[width=0.2\linewidth, height=9mm]{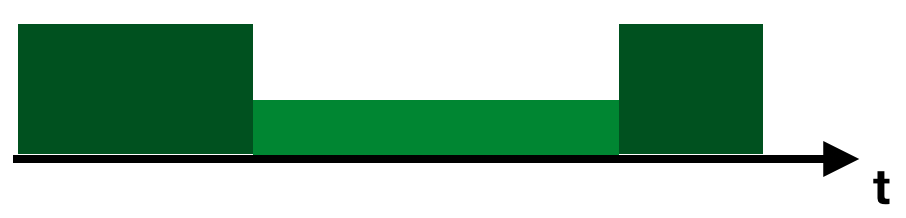}}\\
 \cline{2-7}
    & 1/2 &3/8 & 1/8 & 1/8 & 3/8 &
    \raisebox{-0.5\totalheight}{\includegraphics[width=0.2\linewidth, height=9mm]{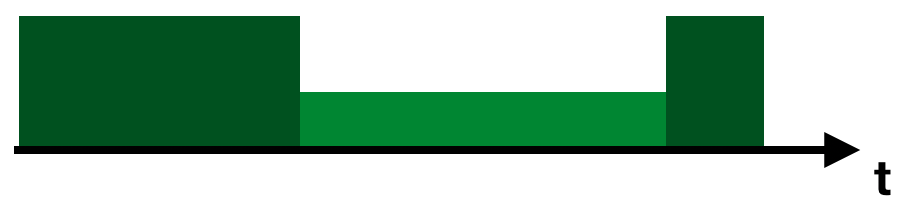}}\\
     \cline{2-7}
    & 3/4 &7/16 & 1/16 & 1/16 & 7/16 &\raisebox{-0.5\totalheight}{\includegraphics[width=0.2\linewidth, height=9mm]{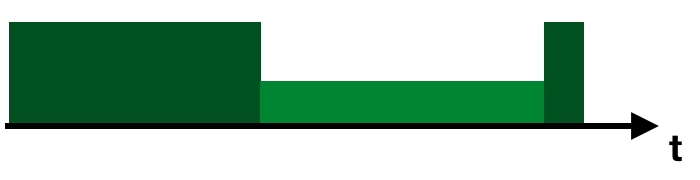}}\\
     \cline{2-7}
    & 1 &1/2 & 0 & 0 & 1/2 &\raisebox{-0.5\totalheight}{\includegraphics[width=0.2\linewidth, height=9mm]{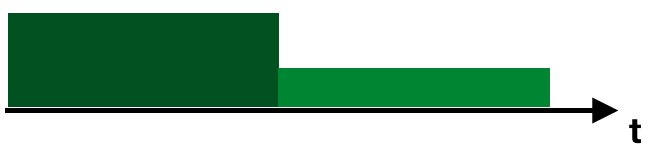}}\\
    \hline
    \hline
    \multirow{4}{*}{3/8}&1/5&3/16 & 3/16 & 3/16 & 7/16 &\raisebox{-0.5\totalheight}{\includegraphics[width=0.2\linewidth, height=18mm]{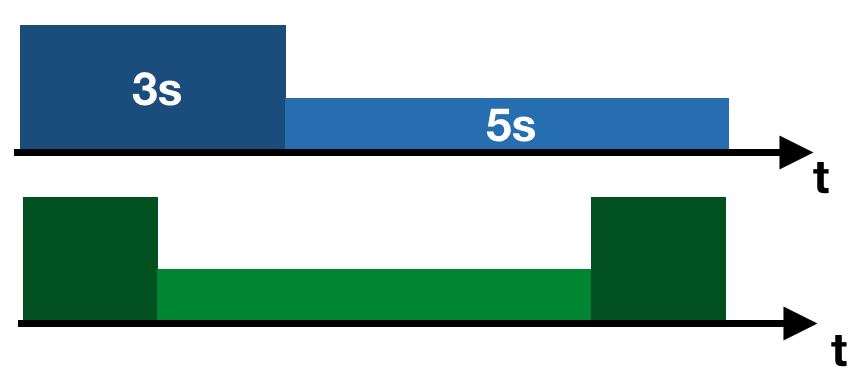}}\\
   \cline{2-7}
    & 7/15 &1/4 & 1/8 & 1/8 & 1/2 &\raisebox{-0.5\totalheight}{\includegraphics[width=0.2\linewidth, height=9mm]{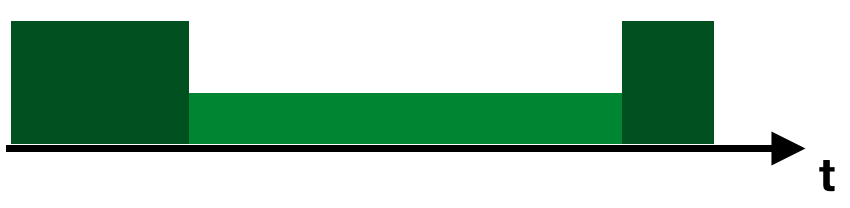}}\\
     \cline{2-7}
    & 11/15 &5/16 & 1/16 & 1/16 & 9/16 &\raisebox{-0.5\totalheight}{\includegraphics[width=0.2\linewidth, height=9mm]{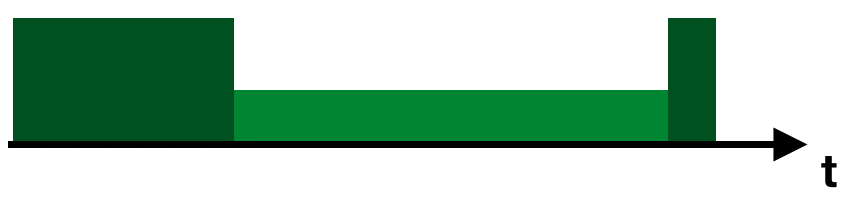}}\\
     \cline{2-7}
    & 1 &3/8 & 0 & 0 & 5/8 &\raisebox{-0.5\totalheight}{\includegraphics[width=0.2\linewidth, height=9mm]{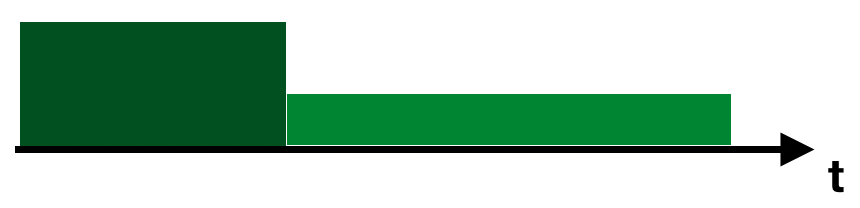}}\\
    \hline
    \hline
     \multirow{3}{*}{1/4}&1/3&1/8 & 1/8 & 1/8 & 5/8 &\raisebox{-0.5\totalheight}{\includegraphics[width=0.2\linewidth, height=18mm]{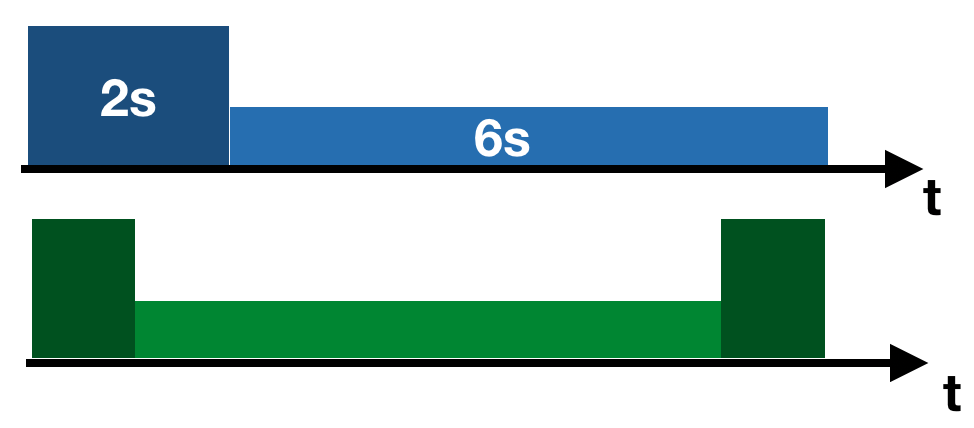}}\\
   \cline{2-7}
    & 2/3 &3/16 & 1/16 & 1/16 & 11/16 &\raisebox{-0.5\totalheight}{\includegraphics[width=0.2\linewidth, height=9mm]{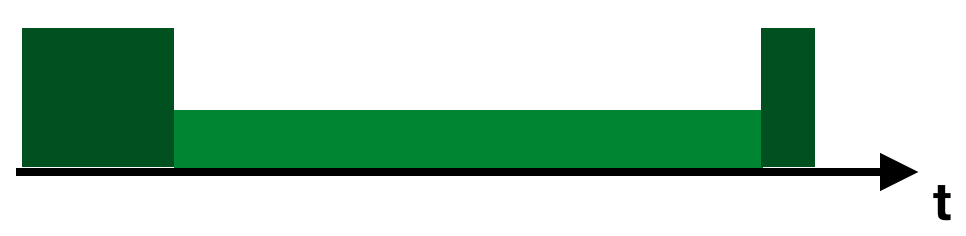}}\\
     \cline{2-7}
    & 1 &1/4 & 0 & 0 & 3/4 &\raisebox{-0.5\totalheight}{\includegraphics[width=0.2\linewidth, height=9mm]{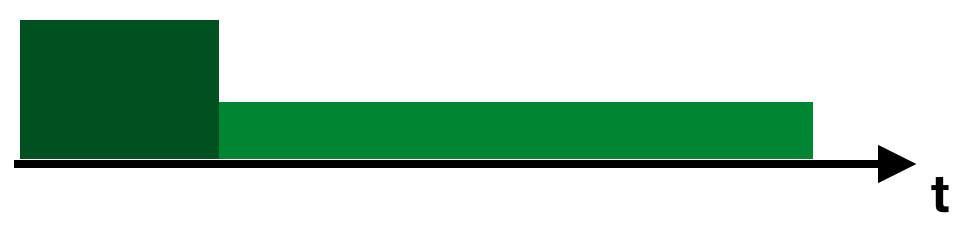}}\\
     \hline
    \hline
    \multirow{2}{*}{1/8}&3/7&1/16 & 1/16 & 1/16 & 13/16 &\raisebox{-0.5\totalheight}{\includegraphics[width=0.2\linewidth, height=18mm]{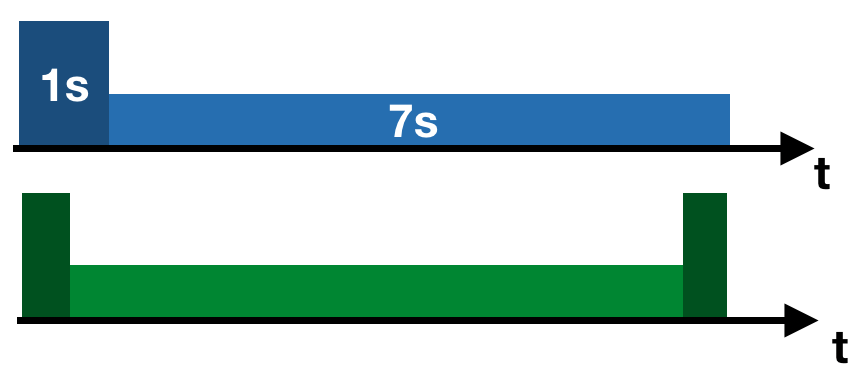}}\\
   \cline{2-7}
    & 1 &1/8 & 0 & 0 & 7/8 &\raisebox{-0.5\totalheight}{\includegraphics[width=0.2\linewidth, height=9mm]{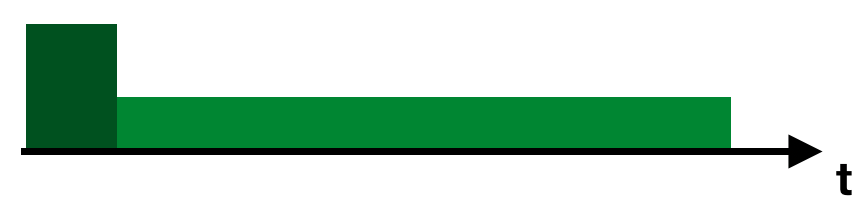}}\\
         \hline
\end{tabular}
\end{center}
\caption{{\it Parameters of the channel.} The coefficients in \eqref{eq:canale} are reported for different choices of $p$ and $\mu$, along with a pictorial representation of the sequences of the voltages to channel 1 (blue) and channel 2 (green).}
\label{tab:multicol}
\end{table*}


Here we report the measured values of the quantities in \eqref{eq:bellob}, pertaining to the operators $X_{i,j}=\sigma_i\otimes\sigma_j$,with $i,j=X,Y,Z$. Theoretical predictions for the ideal channel in Eq. \eqref{eq:canale} are:
\begin{equation}
\left(
    \begin{matrix}
    1 && -1+2p && 1-2p\\
    -1+2p && 1-4p(1-p)(1-\mu) && -1+4p(1-p)(1-\mu)\\
    1-2p && -1+4p(1-p)(1-\mu) && 1-4p(1-p)(1-\mu)
 \end{matrix}
    \right)
\end{equation}
where the $(i,j)$-element in the matrix refers
to $\sigma_i\otimes\sigma_j$, with the index taken in the same order as above.  

The recorded values are as follows (errors in brackets):

$p=1/2,\,\mu=0$
\begin{equation*}
\left(
    \begin{matrix}
    0.9687(5) && 0.020(2) && 0.008(2)\\
0.002(2) && 0.000(2) && 0.003(2)\\
 0.006(2) &&  -0.002(2)&& 0.002(2)\\    \end{matrix}
    \right)
\end{equation*}

$p=1/2,\,\mu=1/4$
\begin{equation*}
\left(
    \begin{matrix}
    0.9685(5) && 0.012(2) && 0.009(2)\\
0.007(2) && 0.243(2) && -0.243(2)\\
 0.006(2) &&  -0.240(2)&& 0.246(2)\\     \end{matrix}
    \right)
\end{equation*}

$p=1/2,\,\mu=1/2$
\begin{equation*}
\left(
    \begin{matrix}
    0.9683(4) && 0.009(2) && 0.013(2)\\
0.007(2) && 0.483(2) && -0.487(2)\\
 0.011(2) &&  -0.483(2)&& 0.487(2)\\     \end{matrix}
    \right)
\end{equation*}

$p=1/2,\,\mu=3/4$
\begin{equation*}
\left(
    \begin{matrix}
    0.9680(5) && 0.011(2) && 0.015(2)\\
0.009(2) && 0.721(2) && -0.733(2)\\
 0.007(2) &&  -0.726(2)&& 0.732(2)\\     \end{matrix}
    \right)
\end{equation*}

$p=1/2,\,\mu=1$
\begin{equation*}
\left(
    \begin{matrix}
    0.9686(4) && 0.014(2) && 0.013(2)\\
0.002(2) && 0.9640(4) && -0.9791(4)\\
 0.009(2) &&  -0.9653(4)&& 0.9720(5)\\     \end{matrix}
    \right)
\end{equation*}

$p=3/8,\,\mu=1/5$
\begin{equation*}
\left(
    \begin{matrix}
    0.9688(6) && -0.224(2) && 0.253(2)\\
-0.236(2) && 0.237(2) && -0.245(2)\\
 0.248(2) &&  -0.239(2)&& 0.244(2)\\     \end{matrix}
    \right)
\end{equation*}

$p=3/8,\,\mu=7/15$
\begin{equation*}
\left(
    \begin{matrix}
    0.9695(5) && -0.225(2) && 0.253(2)\\
-0.237(2) && 0.477(2) && -0.490(2)\\
 0.249(2) &&  -0.479(2)&& 0.499(2)\\     \end{matrix}
    \right)
\end{equation*}

$p=3/8,\,\mu=11/15$
\begin{equation*}
\left(
    \begin{matrix}
    0.9695(5) && -0.225(2) && 0.247(2)\\
-0.237(2) && 0.718(2) && -0.734(2)\\
 0.247(2) &&  -0.720(2)&& 0.731(2)\\     \end{matrix}
    \right)
\end{equation*}

$p=3/8,\,\mu=1$
\begin{equation*}
\left(
    \begin{matrix}
    0.9686(5) && -0.230(2) && 0.254(2)\\
-0.238(2) && 0.9600(6) && -0.9784(4)\\
 0.247(2) &&  -0.9614(6)&& 0.9731(6)\\     \end{matrix}
    \right)
\end{equation*}

$p=1/4,\,\mu=1/3$
\begin{equation*}
\left(
    \begin{matrix}
    0.9683(5) && -0.468(2) && 0.494(2)\\
-0.484(2) && 0.473(2) && -0.486(2)\\
 0.489(2) &&  -0.476(2)&& 0.485(2)\\     \end{matrix}
    \right)
\end{equation*}

$p=1/4,\,\mu=2/3$
\begin{equation*}
\left(
    \begin{matrix}
    0.9681(6) && -0.469(2) && 0.496(2)\\
-0.487(2) && 0.711(2) && -0.732(2)\\
 0.489(2) &&  -0.713(2)&& 0.734(2)\\     \end{matrix}
    \right)
\end{equation*}

$p=1/4,\,\mu=1$
\begin{equation*}
\left(
    \begin{matrix}
    0.9686(5) && -0.468(2) && 0.495(2)\\
-0.483(2) && 0.9528(5) && -0.9764(4)\\
 0.490(2) &&  -0.9574(6)&& 0.9744(6)\\     \end{matrix}
    \right)
\end{equation*}

$p=1/8,\,\mu=3/7$
\begin{equation*}
\left(
    \begin{matrix}
    0.9693(5) && -0.704(2) && 0.734(2)\\
-0.722(2) && 0.708(2) && -0.732(2)\\
 0.729(2) &&  -0.710(2)&& 0.736(2)\\     \end{matrix}
    \right)
\end{equation*}

$p=1/8,\,\mu=1$
\begin{equation*}
\left(
    \begin{matrix}
    0.9686(5) && -0.707(2) && 0.738(2)\\
-0.719(2) && 0.9478(5) && -0.9755(4)\\
 0.729(2) &&  -0.9529(6)&& 0.9756(4)\\     \end{matrix}
    \right)
\end{equation*}






\end{document}